\documentclass[conference]{IEEEtran}
\IEEEoverridecommandlockouts
\renewcommand{\mathrm}[1]{\text{\scriptsize #1}}

\ifCLASSINFOpdf
  \usepackage[pdftex]{graphicx}
  \graphicspath{{../figures/}}
  \DeclareGraphicsExtensions{.pdf,.jpeg,.png}
\else
\fi
\usepackage[font=footnotesize, skip=1pt]{caption}  % Put this in your preamble

\usepackage{amsmath}
\DeclareMathOperator{\sign}{sign}
\usepackage{amssymb}
\usepackage{mathtools}

\usepackage{xcolor}

\bstctlcite{IEEEexample:BSTcontrol}
\newcounter{MYtempeqncnt}

\begin{document}
%\title{Stochastic modeling of beam management in \note{mmWave} 3GPP NR}
%\title{A stochastic model for beam alignment in mmWave 3GPP NR}
%\title{SSB-based beam management in mmWave NR}


\title{Beam Misalignment in 3GPP mmWave NR
\thanks{This work has received funding from the EU's Horizon Europe research and innovation programme under the Marie Skłodowska-Curie grant agreement No. 101073265 (EWOC). Views and opinions expressed are however those of the authors only and do not necessarily reflect those of the EU. The EU cannot be held responsible for them.}}

\author{\IEEEauthorblockN{Noè~Bernadas~i~Busquets\textsuperscript{1,2}, Xavier~Gelabert\textsuperscript{2}, Bleron~Klaiqi\textsuperscript{2}, Ki~Won~Sung\textsuperscript{1}, Slimane~Ben~Slimane\textsuperscript{1}}
\IEEEauthorblockA{\textit{\textsuperscript{1}Royal Institute of Technology (KTH), Stockholm, Sweden}}
\IEEEauthorblockA{\textit{\textsuperscript{2}Huawei Technologies Sweden AB, Stockholm Research Centre, Kista, Sweden}}
noebib@kth.se, \{bleron.klaiqi, xavier.gelabert\}@huawei.com, \{sungkw, slimane\}@kth.se}

\maketitle

\bstctlcite{IEEEexample:BSTcontrol}
\begin{abstract}
This paper presents an analytical framework for evaluating beam misalignment in 3GPP mmWave NR systems implementing analog beamforming. Our approach captures the interaction between user mobility, beam sweeping mechanisms, and deployment configurations, focusing on long-term average performance metrics. Specifically, we model the beam misalignment rates at both the base station (BS) and user equipment (UE) as Poisson processes and derive expressions for the expected misalignment duration, misalignment fraction, and overall beamforming gain. The framework accounts for practical constraints in NR such as Synchronization Signal Blocks (SSB) periodicity, TDD frame structures, and SSB overhead. Through numerical evaluation based on 3GPP mmWave parameters, we identify key trade-offs between beam counts, user mobility, and SSB timing, providing actionable design insights for robust and efficient beam management in future high-frequency networks.
\end{abstract}


% \fi
%\IEEEpeerreviewmaketitle

\section{Introduction}
% for IEEE journal papers produced under \LaTeX\ using
% IEEEtran.cls version 1.8b and later.

Future wireless systems are expected to support dynamic, dense, and heterogeneous deployments, calling for new techniques allowing reliable connectivity, low-latency communication, and robust mobility management~\cite{vujicic2024}. Among the key enablers, the use of higher frequency bands has been identified to meet the demands of emerging services in next-generation networks~\cite{zugno2025a}. It is well established that cellular communication operating in high frequency bands, e.g., in the FR2 mmWave bands and beyond, experiences higher power loss yielding smaller coverage areas. This effect can be partially mitigated by deploying multi-antenna arrays, where the superimposition of signals from different antennas, with controlled amplitudes and phases (\emph{weights}), yields a narrow-width beam-like radiation pattern in a certain direction. The process of adjusting these amplitudes and phases to steer and shape the beam is known as \emph{beamforming} (BF) \cite{emilbjornsonIntroductionMultipleAntenna2024}.

Two BF methods are typically considered: (1) static weighting, where a fixed set of predefined weights generates a corresponding number of static beams, each with fixed beamwidth and direction; and (2) dynamic weighting, where weights are computed adaptively based on real-time channel conditions. In this work, we assume static analog BF, performed in the RF domain, being able to support one active beam per transmission time interval. Due to its hardware efficiency and power savings, analog BF has been widely adopted in mmWave systems \cite{emilbjornsonIntroductionMultipleAntenna2024} and the procedure specified in 3GPP \cite{3gpp_ts_38213}.

With this in mind, the primary objective of beam management is to establish and maintain \emph{aligned} beam pairs, i.e. the main lobes of the transmitting and receiving antenna arrays are directed toward each other. This ensures a robust wireless link between the base station (BS) and the user equipment (UE). This is realized via the \emph{serving beam selection} procedure, composed of \emph{beam sweeping}, \emph{measurement}, \emph{reporting}, and \emph{indication} \cite{3gpp_tr_38912}. As UEs move, beam alignment must be continuously adjusted. In 3GPP NR, wide beam alignment, also referred to as \emph{Process P1}, is achieved through the periodic transmission of Synchronization Signal Blocks (SSBs). A BS transmits $N_{\mathrm{beam}}^{\mathrm{BS}}$ SSB beams across a sweeping schedule, and the UE measures these over $N_{\mathrm{beam}}^{\mathrm{UE}}$ receive beams. Based on the measurements, a beam report is sent to the BS, which indicates the selected serving beam to the UE. A more detailed description of beam management can be found in, e.g. \cite{3gpp_tr_38802,giordani2019}. Narrow beam refinement procedures (Processes P2 and P3) are also implemented but are not addressed in this work.

Prior work on beam management has largely focused on simulation-based studies or heuristic design of sweeping strategies \cite{giordani2019}. Existing analytical models typically simplify mobility dynamics or overlook key timing constraints imposed by 3GPP standards. Some studies address beam selection delay \cite{beamdelay}, while others propose optimization-based approaches for tracking and reporting beams \cite{maggi_beam, zhang_beam}. However, few works analytically characterize misalignment time distributions or their coupling with system-level parameters such as SSB periodicity and TDD configuration. This paper closes that gap by integrating geometric, timing, and beam control models into a unified performance analysis framework.

This paper presents a novel analytical framework for quantifying beam misalignment in mmWave systems using static analog BF. In contrast to prior works that often abstract away the impact of beam sweeping dynamics and mobility, our model integrates a stochastic geometry-based mobility process with a 3GPP-compliant SSB sweeping timing model. We derive closed-form expressions for misalignment probabilities, expected misalignment durations, average BF gain, and SSB-induced overhead across varying deployment and mobility scenarios. Numerical evaluations under realistic NR configurations validate the framework and highlight key trade-offs, offering design insights for robust beam management in mmWave networks.

The remainder of the paper is organized as follows. Section~\ref{sec:system_model} presents the system model, including deployment and mobility assumptions, the 3GPP-compliant SSB sweeping timing structure, and the formulation of misalignment events and durations. Section~\ref{sec:numerical_evaluation} provides numerical evaluations and key insights. Finally, Section~\ref{sec:conclusion} concludes the paper.

\section{System Model}
\label{sec:system_model}
%\subsection{Deployment and beam modeling}
We consider a stochastic geometry-based network deployment where the locations of BSs are modelled as a homogeneous Poisson Point Process (PPP) $\Phi \!\!\subset\!\! \mathbb{R}^2$ with intensity $\lambda$ BSs/m\textsuperscript{2}. The relation with the average distance between BSs, or inter-site distance (ISD), is $\lambda \!\!=\!\! 4/(\pi d_{\mathrm{ISD}})$ \cite{kalamkar2022}.

% \begin{equation}
%     \lambda = \frac{4}{\pi \, d_{\mathrm{ISD}}}.
% \end{equation}
The BF model proposed by \cite{kalamkar2022} is assumed where each BS employs $N_{\mathrm{beam}}^{\mathrm{BS}}$ directional beams with predefined beamwidths $\varphi^{\mathrm{BS}} =2\pi/N_{\mathrm{beam}}^{\mathrm{BS}}$. The beamformed antenna pattern at an angle $\psi$ measured from the beam's boresight direction is given by:
\begin{equation}
\label{eq:beam_gain}
G^{\mathrm{BS}}(\psi) =
\begin{cases} 
G_{\mathrm{main}}^{\mathrm{BS}}=N_{\mathrm{beam}}^{\mathrm{BS}}, & \text{if } |\psi| \leq \varphi^{\mathrm{BS}} / 2, \\
G_{\mathrm{side}}^{\mathrm{BS}}=1/N_{\mathrm{beam}}^{\mathrm{BS}}, & \text{otherwise},
\end{cases}
\end{equation}
where the main lobe gain $G_{\mathrm{main}}^{\mathrm{BS}}$ scales with the number of beams, while the sidelobe gain $G_{\mathrm{side}}^{\mathrm{BS}}$ scales inversely.

Mobile users (UEs) traverse the network with a velocity $v$, following a straight line mobility model with random orientation.  Similarly, UEs employs a BF model with $N_{\mathrm{beam}}^{\mathrm{UE}}$ beams, each with beamwidth $\varphi^{\mathrm{UE}}$. The corresponding antenna gain at an angle $\psi$, $G^{\mathrm{UE}}(\psi)$, mirrors that of (\ref{eq:beam_gain}) with $G_{\mathrm{main}}^{\mathrm{UE}} = N_{\mathrm{beam}}^{\mathrm{UE}}$, 
 and $G_{\mathrm{side}}^{\mathrm{UE}}=1/N_{\mathrm{beam}}^{\mathrm{UE}}$.

The impact of user mobility on beam alignment is characterized by the probability of beam misalignment, which depends on the UE speed $v$, the number of beams $N_{\mathrm{beam}}^{(i)}$, with $i\in\{\text{\small BS},\text{\small UE}\}$, and network topology (via  $\lambda$). The expression for the beam misalignment probability during an interval $\tau$ follows a Poisson distribution and is given by \cite{kalamkar2022}:
\begin{equation}
\label{eq:prob_beam_missalign}
\begin{aligned}
p_{\mathrm{M}}^{(i)} &= 1 - \exp \left( -\frac{N_{\mathrm{beam}}^{(i)} \sqrt{\lambda} v \tau}{\pi} \right) \triangleq 1- e^{-\beta^{(i)} \tau}   ,    
\end{aligned}
\end{equation}
where $p_{\mathrm{M}}^{(i)}$, $i\in\{\text{\small BS},\text{\small UE}\}$, indicates the probability the UE (BS) is outside the main lobe of the serving BS (UE) beam within an observation period $\tau$. $\beta^{(i)}$ are the misalignment rates.

\subsection{Beam management timing model}
\label{subsec:beam_management_timing}
%\note{[Here introduce the SSB-based beam sweeping, its duration and the obtained sweeping times. For ICTON, because of 4-page limitation, it should be mainly descriptive, but one could still include the sweeping time figure, and describe those timing aspects that are necessary for the next subsection.]}

%The duration of SSB-based beam sweeping (henceforth SSB beam sweeping) depends on the number of beams at both the BS and UE, as well as the time required for measurement processing and reporting. In this respect, \cite{giordani2019} presents an approximate expression for SSB beam sweeping duration. However, it relies on several simplifying assumptions, such as fixed SSB starting symbol positions across slots and the omission of non-contiguous SSB slot allocations due to uplink (UL) and downlink (DL) slot configurations in a given time-division duplexing (TDD) frame structure. In this work, we relax these assumptions and consider a more general SSB scheduling model as per 3GPP specifications \cite{3gpp_ts_38213, 3gpp_ts_38331}. 
The duration of SSB-based beam sweeping depends on the number of BS and UE beams, as well as the time for measurement processing and reporting. While \cite{giordani2019} provides an approximate expression, it assumes fixed SSB symbol positions and ignores non-contiguous slot allocations due to uplink (UL) and downlink (DL) slot configurations in a given TDD frame structure. In this work, we relax these assumptions and adopt a more general 3GPP-compliant SSB scheduling model \cite{3gpp_ts_38213, 3gpp_ts_38331}, which is presented next.

%\cite{3gpp_tr_38912}
Each SSB occupies $N_{\mathrm{SSB}}^{\mathrm{symb}}=4$ symbols in the time domain and spread over 240 subcarriers in the frequency domain. One or more SSBs form an SS burst, and multiple SS bursts compose an SS burst set, whose size is upper-bounded by $L_{\mathrm{SSB}}$. The number of SSBs per slot is  $N_{\mathrm{SSB}}^{\mathrm{slot}}=\{1,2\}$ \cite{3gpp_ts_38133}. SS burst sets are transmitted periodically with $\tau_{\mathrm{SS}}=\{5,10,20,40,80,160\}$ms, where $\tau_{\mathrm{SS}}=20$ms is commonly used during initial access procedure. Each SS burst set occupies a fixed $T_{\mathrm{SS}}=5$ ms window, located either in the first or second half of a 10 ms radio frame (first half assumed). The maximum number of candidate SSBs ($L_{\mathrm{SSB}}$) and their start symbols $l_{\mathrm{SSB}}(i)$ (relative to the first symbol of the half-frame) depend on carrier frequency and subcarrier spacing as specified by \cite{3gpp_ts_38213}. Table \ref{tab:max_ssb_table} lists the calculated values of $L_{\mathrm{SSB}}$ as defined by 3GPP in \cite{3gpp_ts_38213} for subcarrier spacings of 120, 480 and 960 kHz (cases D, F and G). Since these values assume no specific TDD framestructure, we also compute the effective number of SSBs per SS burst set for TDD configurations (a) and (b) as per \cite{ecc_rec_2003}. Details are given in Appendices~\ref{appA} and \ref{appB}.

\begin{table}[!t]
\caption{Maximum number of SSBs within a SS burst set for different SCSs and TDD framestructure options.}
\label{tab:max_ssb_table}
\centering
\resizebox{\columnwidth}{!}{%
\begin{tabular}{|c|c|c|c|}
\hline
\textbf{Case - SCS} & \textbf{3GPP \cite{3gpp_ts_38213}} & \textbf{TDD Option (a)} & \textbf{TDD Option (b)}  \\
\hline
D - 120 kHz & 
$L_{\mathrm{SSB}}^\mathrm{\mathrm{D}}=64$
 & 
 $L_{\mathrm{SSB,eff}}^{\mathrm{(a),D}}=\{52,56\}$
 &    $L_{\mathrm{SSB,eff}}^{\mathrm{(b),D}}=\{50,52\}$\\
\hline
F - 480 kHz & $L_{\mathrm{SSB}}^\mathrm{F}=64$ & $L_{\mathrm{SSB,eff}}^{\mathrm{(a),F}}=64$ & $L_{\mathrm{SSB,eff}}^{\mathrm{(b),F}}=64$   \\
\hline
G - 960 kHz  & $L_{\mathrm{SSB}}^\mathrm{G}=64$ & $L_{\mathrm{SSB,eff}}^{\mathrm{(a),G}}=64$ & $L_{\mathrm{SSB,eff}}^{\mathrm{(b),G}}=64$   \\
\hline
\end{tabular}
}
\end{table}

The number of requested beamformed SSB transmissions is given by the product \( N_{\mathrm{SSB}}^{\mathrm{req}} = N_{\mathrm{beam}}^{\mathrm{BS}} \cdot N_{\mathrm{beam}}^{\mathrm{UE}} \), which determines all possible beam direction combinations between the BS and the UE. We define the \emph{sweep time} as the time required to complete all beam sweeps. The number of requested SSBs, $N_{\mathrm{SSB}}^{\mathrm{req}}$, may exceed the available number of SSBs in a given SS burst set, denoted by  $L_{\mathrm{SSB,eff}}^{p,\mu}$ (see Table \ref{tab:max_ssb_table}). In this case, the residual SSBs will be transmitted in the subsequent SS burst set. Therefore, the total sweep time, \( T_{\mathrm{sweep}} \), is the sum of the time for fully occupied (complete) SS burst sets and, if any, the residual contribution from the remaining SSBs, which occupy only a fraction of the SS burst set. Specifically, 
\begin{equation}
T_{\mathrm{sweep}} = T_{\mathrm{sweep,c}} + T_{\mathrm{sweep,r}},  
\end{equation}
where $T_{\mathrm{sweep,c}}$ represents the time for complete SS burst sets and $T_{\mathrm{sweep,r}}$ represents the time for the residual SSBs. The exact calculation of $T_{\mathrm{sweep,c}}$ and $T_{\mathrm{sweep,r}}$ is provided in appendix \ref{appC}. We assume that a sweep starts always at the beginning of an SS burst set. The sweep periodicity $\tau_{\mathrm{sweep}}$ is then:
\begin{equation}
\tau_{\mathrm{sweep}} = \left\lceil \frac{T_{\mathrm{sweep}}}{\tau_{\mathrm{SS}}} \right\rceil \cdot \tau_{\mathrm{SS}} \triangleq N_{\mathrm{sweep}} \cdot \tau_{\mathrm{SS}},    
\end{equation}
where $N_{\mathrm{sweep}}$ has been introduced to denote the number of SS burst sets required to complete the beam sweeping process. 

%Once beam sweeping is completed, there is a time needed to perform measurements, report said measurements to the network and indicate possible beam changes. This time we call processing time $T_{\mathrm{proc}}$ and will be assumed deterministic. 

\begin{figure}[t]
    \centering
    \includegraphics[width=8cm]{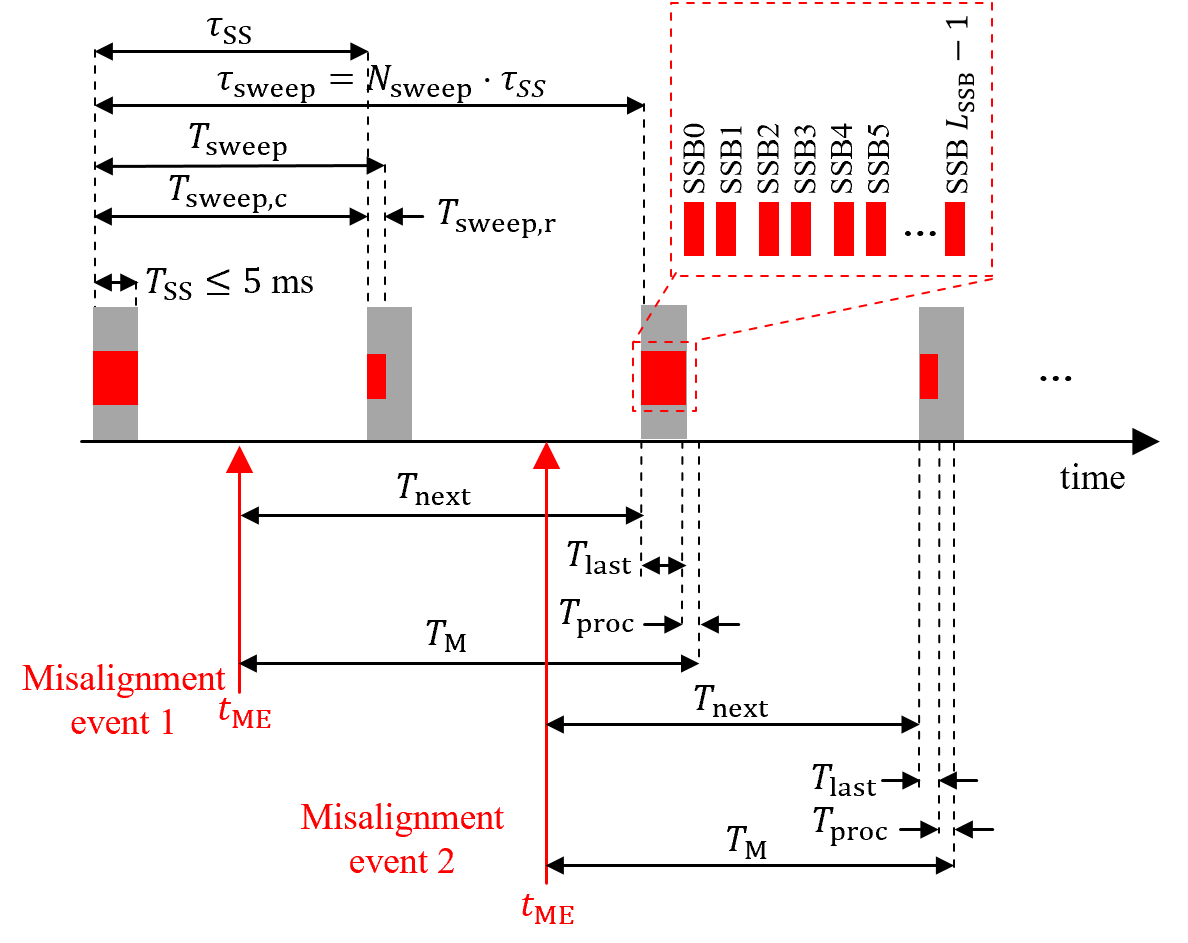}
    \caption{The considered SSB timing model showing two representative missalignment events.}
    \label{fig:timing_model}
\end{figure}

Within this time frame, a misalignment event can occur at random times $t_{\mathrm{ME}}$, assumed to be uniformly distributed over interval $[0, \tau_{\mathrm{sweep}}]$. Once the misalignment event occurs, the total time during which the misalignment persists is given by:
\begin{equation}
\label{eq:missalignment_duration}
T_{\mathrm{M}} = T_{\mathrm{next}} + T_{\mathrm{last}} + T_{\mathrm{proc}},
\end{equation}
where $T_{\mathrm{next}}$ is the time between the misalignment event and the start of the next sweep period, $T_{\mathrm{last}}$ represents the time needed for the last SSB transmissions within the sweep, and $T_{\mathrm{proc}}$ is the processing time needed for measurement and reporting. Fig. \ref{fig:timing_model} illustrates the timing model with the defined time parameters showing two representative misalignment events. 

\begin{figure}
    \centering
    \includegraphics[width=8cm]{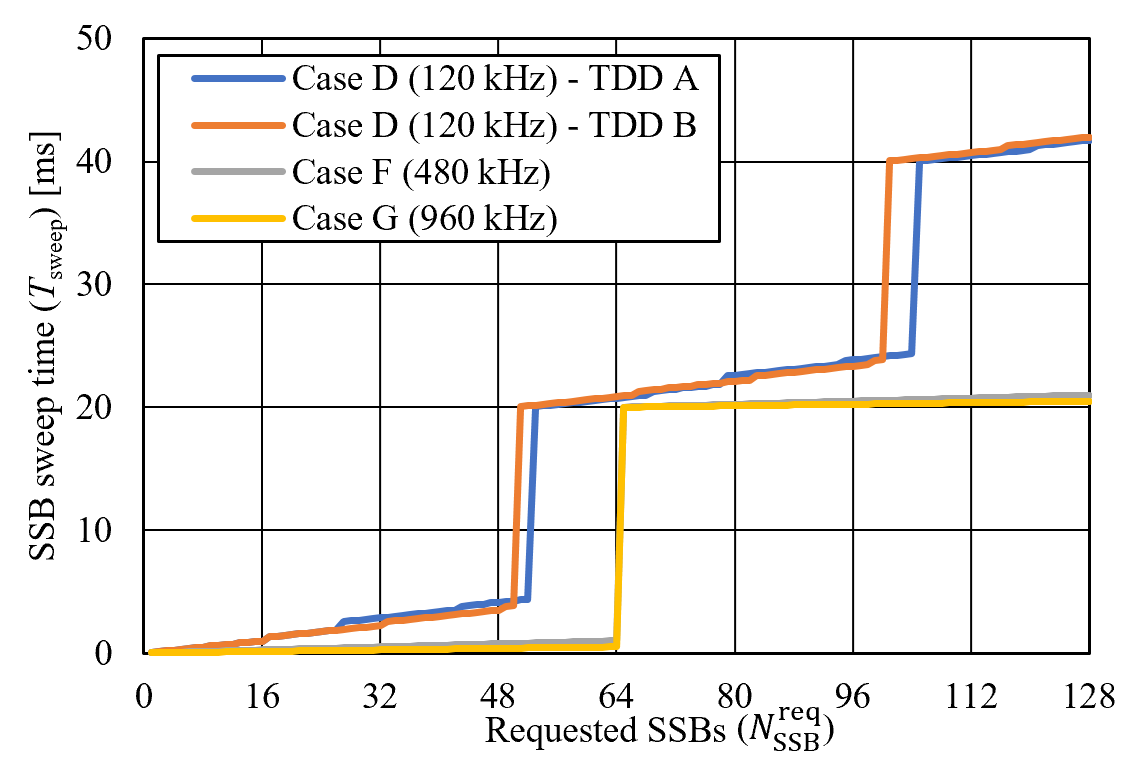}
    \caption{Sweep duration ($T_{\mathrm{sweep}}$) against the number of requested SSBs ($N_{\mathrm{SSB}}^{\mathrm{req}}$) for different TDD frame configurations and subcarrier spacings.}
    \label{fig:sweep_duration}
\end{figure}

In Fig. \ref{fig:sweep_duration}, we show for a representative number of deployment cases (see Table \ref{tab:max_ssb_table}), the sweep time duration ($T_{\mathrm{sweep}}$) against the number of requested SSBs ($N_{\mathrm{SSB}}^{\mathrm{req}}$). When the requested number of SSBs exceeds the number of supported SSBs per SS burst set in a particular setup, a step-wise increase in sweeping time is observed as a result of having to wait until the next SS burst period to transmit the remaining SSBs.  

\subsection{Misalignment duration modelling}

We consider independent beam misalignment processes at the BS and UE, as characterized by the Poisson process in (\ref{eq:prob_beam_missalign}), by which the rate $\beta^{(i)}$, $i \in \{\text{\small BS,UE}\}$, represent the average number of misalignment events per second. 

Assuming a uniform distribution of the misalignment event time $t_\mathrm{ME}$ over a beam sweeping period $\tau_\mathrm{sweep}$, we can write
\begin{equation}
\mathbb{E}[T_\mathrm{next}^{(i)}] = \tau_\mathrm{sweep}/{2}.    
\end{equation}

For the average time needed to transmit the remaining SSBs in the ongoing sweep cycle, $\mathbb{E}[T_\mathrm{last}^{(i)}]$, two different cases are identified: (a) the remaining SSBs \emph{completely} occupies the SS burst duration ($T_{\mathrm{SS}}$) and (b) the remaining SSBs \emph{partially} occupies the burst duration. Assuming a uniform distribution of the misalignment event times, the probabilities associated with these two cases are $1-1/N_\mathrm{sweep}$ and $1/N_\mathrm{sweep}$ respectively. For each of these cases, the duration is $T_{\mathrm{SS}}$ and $T_\mathrm{sweep,r}$, as can be seen in Fig. \ref{fig:timing_model}. Consequently, the average duration of transmitting the ramainig SSBs is given by:
\begin{equation}
    \mathbb{E}[T_\mathrm{last}^{(i)}]= (1-\frac{1}{N_\mathrm{sweep}})\cdot T_{\mathrm{SS}} + \frac{1}{N_\mathrm{sweep}}\cdot T_\mathrm{sweep,r}  
\end{equation}
Finally, the processing time is considered to be determininstic, so $\mathbb{E}[T_\mathrm{proc}^{(i))}]=T_\mathrm{proc}$. Consequently, from (\ref{eq:missalignment_duration}), the average missalignment time can be expressed as:
\begin{equation}
\label{eq:avg_misalignment_duration}
\begin{aligned}
\mathbb{E}[T_\mathrm{M}^{(i)}] & = \mathbb{E}[T_\mathrm{next}^{(i)}] + \mathbb{E}[T_\mathrm{last}^{(i)}] + \mathbb{E}[T_\mathrm{proc}^{(i)}]  = \frac{\tau_\mathrm{sweep}}{2} \\
& + (1-\frac{1}{N_\mathrm{sweep}})\cdot T_{\mathrm{SS}}  + \frac{1}{N_\mathrm{sweep}}\cdot T_\mathrm{sweep,r} + T_\mathrm{proc}   .
\end{aligned}
\end{equation}
%By inspecting (\ref{eq:avg_misalignment_duration}), the average duration of a single misalignment event, whether due to a BS beam or a UE beam, is determined solely by the timing of the beam management procedure (e.g., sweeping interval and processing time) and not by which end (BS or UE) became misaligned. Therefore, under symmetric beam management mechanisms, the expected duration of BS and UE misalignment events can be considered equal, i.e. $\mathbb{E}[T_{\mathrm{M}}^{\mathrm{BS}}] = \mathbb{E}[T_{\mathrm{M}}^{\mathrm{UE}}]$. This will be the case considered hereon.
From (\ref{eq:avg_misalignment_duration}), the average duration of a misalignment event, whether caused by the BS or UE, depends only on the beam management timing (e.g., sweeping interval and processing delay), not on which side is misaligned. Thus, under symmetric beam management, we will assume $\mathbb{E}[T_{\mathrm{M}}^{\mathrm{BS}}] = \mathbb{E}[T_{\mathrm{M}}^{\mathrm{UE}}]$.

The duration of misalignment events given by (\ref{eq:avg_misalignment_duration}) accounts for either BS or UE missalignment events. However, misalignment events at the BS and UE can overlap in time. We assume the two misalignment processes are independent Poisson processes, as per (\ref{eq:prob_beam_missalign}).

Let us define an \emph{overlapping misalignment} as the situation in which a second misalignment event (e.g., at the UE) begins while the first misalignment (e.g., at the BS) is still unresolved. The resulting missalignment is not simply $T_\mathrm{M}^{\mathrm{BS}} + T_\mathrm{M}^{\mathrm{UE}}$, since their missalignment windows can overlapp. Assume the BS becomes misaligned at $t=0$, and remains missaligned for a time $T_\mathrm{M}^{\mathrm{BS}}$. Now assume the UE becomes misaligned at a random time $t=\delta$ during that BS misalignment period, i.e. $0\le \delta \le T_\mathrm{M}^{\mathrm{BS}}$. Considering $\delta$ uniformly distributed between the interval $[0,T_\mathrm{M}^{\mathrm{BS}}]$, the total overlapping misalignment time is also random with $T_\mathrm{M}^{\mathrm{BU}}=\max\left(T_\mathrm{M}^{\mathrm{BS}},\delta + T_\mathrm{M}^{\mathrm{UE}} \right)$. We now compute the average overlapping duration as:
% \begin{equation}
% \label{eq:overlap_duration}
% \begin{aligned}
%     \mathbb{E}[T_\mathrm{M}^{\mathrm{BU}}]& =\mathbb{E}[\max\left(T_\mathrm{M}^{\mathrm{BS}},\delta + T_\mathrm{M}^{\mathrm{UE}} \right)] \\ 
%     & = \frac{1}{T_\mathrm{M}^{\mathrm{BS}}}\int_0^{T_\mathrm{M}^{\mathrm{BS}}}\max\left(T_\mathrm{M}^{\mathrm{BS}},\delta + T_\mathrm{M}^{\mathrm{UE}} \right) d\delta\\
%     & = \frac{1}{T_\mathrm{M}^{\mathrm{BS}}} \left( \right. \int_0^{T_\mathrm{M}^{\mathrm{BS}}-T_\mathrm{M}^{\mathrm{UE}}} T_\mathrm{M}^{\mathrm{BS}} d\delta \\ &+ \int_{T_\mathrm{M}^{\mathrm{BS}}-T_\mathrm{M}^{\mathrm{UE}}}^{T_\mathrm{M}^{\mathrm{BS}}} \left(\delta + T_\mathrm{M}^{\mathrm{UE}}\right) d\delta \left. \right) \\
%     &=T_\mathrm{M}^{\mathrm{BS}}+\frac{(T_\mathrm{M}^{\mathrm{UE}})^2}{2T_\mathrm{M}^{\mathrm{BS}}}.
% \end{aligned}
% \end{equation}
\begin{equation}
\label{eq:overlap_duration}
\begin{aligned}
    &\mathbb{E}[T_\mathrm{M}^{\mathrm{BU}}]=\mathbb{E}[\max\left(T_\mathrm{M}^{\mathrm{BS}},\delta + T_\mathrm{M}^{\mathrm{UE}} \right)] \\ 
    &= \frac{1}{T_\mathrm{M}^{\mathrm{BS}}}\int_0^{T_\mathrm{M}^{\mathrm{BS}}}\!\!\!\!\max\left(T_\mathrm{M}^{\mathrm{BS}},\delta + T_\mathrm{M}^{\mathrm{UE}} \right) d\delta=T_\mathrm{M}^{\mathrm{BS}}+\frac{(T_\mathrm{M}^{\mathrm{UE}})^2}{2T_\mathrm{M}^{\mathrm{BS}}}.
\end{aligned}
\end{equation}
If we now consider the misalignment durations $T_\mathrm{M}^{\mathrm{BS}}$ and $T_\mathrm{M}^{\mathrm{UE}}$ as random variables with means $\mathbb{E}[T_\mathrm{M}^{\mathrm{BS}}]$ and $\mathbb{E}[T_\mathrm{M}^{\mathrm{UE}}]$ respectively, then, assuming independent processes and Jensen’s inequality \cite{kobayashiProbabilityRandomProcesses2012}, we can rewrite (\ref{eq:overlap_duration}) as:
 \begin{equation}
 \label{eq_avg_over2}
        \mathbb{E}[T_\mathrm{M}^{\mathrm{BU}}] \approx \mathbb{E}[T_\mathrm{M}^{\mathrm{BS}}] + \frac{\mathbb{E}[(T_\mathrm{M}^{\mathrm{UE}})^2]}{2\mathbb{E}[T_\mathrm{M}^{\mathrm{BS}}]}
    \end{equation}

The term $\mathbb{E}[x^2]$ in (\ref{eq_avg_over2}) can be expressed as $\mathbb{E}[x^2]=Var(x)+\mathbb{E}[x]^2$, where for a uniform distribution over $[a,b]$ the variance becomes $Var(x) = (b-a)^2/12$ \cite{kobayashiProbabilityRandomProcesses2012}. By using this and by considering $\mathbb{E}[T_\mathrm{M}^{\mathrm{BS}}]=\mathbb{E}[T_\mathrm{M}^{\mathrm{UE}}]$,
\begin{equation}
\label{eq:overlapp_duration}
        \mathbb{E}[T_\mathrm{M}^{(\mathrm{overlapp})}] \triangleq \mathbb{E}[T_\mathrm{M}^{(\mathrm{BU})}] = \frac{5}{3}\mathbb{E}[T_\mathrm{M}^{\mathrm{BS}}]= \frac{5}{3}\mathbb{E}[T_\mathrm{M}^{\mathrm{UE}}].
\end{equation} 

%\subsection{Overall Misalignment}
To characterize the overall impact of beam misalignments on system performance, we define an \emph{average beam misalignment duration} ($\Gamma$), which captures the long-term average misaligned duration. This metric accounts for three types of misalignment scenarios: (i) BS-only misalignment, (ii) UE-only misalignment, and (iii) overlapping misalignments where both BS and UE are misaligned simultaneously. Each case is described by its occurrence probability and expected duration. 

The expected fraction of time that the system is misaligned due to $i\in\{\text{\small BS,UE}\}$ events can be defined by:
\begin{equation}
\label{eq:gamma_BS}
    \gamma_{(i)} = \beta^{(i)} \cdot \mathbb{E}[T_\mathrm{M}^{(i)}],
\end{equation}
% \begin{equation}
% \label{eq:gamma_UE}
%     \gamma_\mathrm{UE} = \beta^\mathrm{UE} \cdot \mathbb{E}[T_\mathrm{M}^{\mathrm{UE}}],
% \end{equation}
with $\beta^{(i)}$ the beam misalignment rate (in events/s), see (\ref{eq:prob_beam_missalign}), and $\mathbb{E}[T_\mathrm{M}^{(i)}]$ the expected misalignment durations, as per (\ref{eq:avg_misalignment_duration}). These expressions follow directly from Little’s formula  \cite{kobayashiProbabilityRandomProcesses2012}, where $\gamma_{(i)}$ denotes the average number of active misalignments, or equivalently, the fraction of time the system is misaligned. For the model to be valid, we require $\gamma_{(i)} \le 1$, ensuring the system can recover before the next misalignment occurs.
    
We now define the \emph{total misalignment fraction} as the fraction of time during which at least one end (BS or UE) is misaligned. This is given by:
\begin{equation}
\label{eq:gamma_total}
    \gamma_\mathrm{total} = \gamma_\mathrm{BS} + \gamma_\mathrm{UE} - \gamma_\mathrm{BS} \cdot \gamma_\mathrm{UE}.
\end{equation}

Equation~\eqref{eq:gamma_total} reflects the inclusion-exclusion principle applied to the probability of overlapping misalignment events. The term $\gamma_\mathrm{BS} + \gamma_\mathrm{UE}$ accounts for the total fraction of time when either the BS or the UE is misaligned. However, since the intervals during which both are misaligned are counted twice in this sum, we subtract the overlap, represented by $\gamma_\mathrm{BS} \cdot \gamma_\mathrm{UE}$, to avoid double-counting. This assumes that the BS and UE misalignment processes are independent, and allows the use of \eqref{eq:gamma_total} as a reliable approximation of the total time the system operates in a misaligned state.

To assess the relative impact of different misalignment types we define the following normalized weights:
\begin{equation}
\begin{aligned}
P_{\mathrm{B}} \!=\! \frac{\gamma_\mathrm{BS} (1\!-\!\gamma_\mathrm{UE})}{\gamma_\mathrm{total}}, 
P_{\mathrm{U}}\! =\! \frac{\gamma_\mathrm{UE} (1\!-\!\gamma_\mathrm{BS})}{\gamma_\mathrm{total}},
P_{\mathrm{BU}} \!=\! \frac{\gamma_\mathrm{BS} \!\cdot\! \gamma_\mathrm{UE}}{\gamma_\mathrm{total}}.
\end{aligned}
\end{equation}

These quantities represent the normalized probabilities that a randomly selected misalignment time interval corresponds to a BS-only misalignment ($P_\mathrm{B}$), a UE-only misalignment ($P_\mathrm{U}$), or a joint misalignment of both BS and UE ($P_\mathrm{BU}$). Their sum is equal to 1 by construction, i.e., $P_{\mathrm{B}} + P_{\mathrm{U}} + P_{\mathrm{BU}} = 1 $.

The derivation assumes that, again, the BS and UE misalignment processes are statistically independent, and that the misalignment durations are sufficiently short and well-separated to treat overlaps probabilistically rather than through detailed time alignment. The terms $\gamma_\mathrm{BS}$ and $\gamma_\mathrm{UE}$ reflect the long-term time fractions of misalignment caused by each side, while $\gamma_\mathrm{total}$ quantifies the total fraction of time during which the system is in a misaligned state, as per Eq.~\eqref{eq:gamma_total}.

The overall misalignment duration is then expressed as the weighted average of the misalignment durations:
\begin{equation}
\label{eq:gamma_total_duration}
\Gamma = P_{\mathrm{B}} \cdot \mathbb{E}[T_\mathrm{M}^{\mathrm{BS}}]
+ P_{\mathrm{U}} \cdot \mathbb{E}[T_\mathrm{M}^{\mathrm{UE}}]
+ P_{\mathrm{BU}} \cdot \mathbb{E}[T_\mathrm{M}^{\mathrm{BU}}].
\end{equation}
where $\mathbb{E}[T_\mathrm{M}^{\mathrm{BU}}]$ is the expected duration of overlapping misalignment defined in (\ref{eq:overlapp_duration}).
In (\ref{eq:gamma_total_duration}) we capture both how frequently misalignments occur and how long they persist, offering a suitable measure of the temporal degradation caused by beam misalignment.

\section{Numerical Evaluation}
\label{sec:numerical_evaluation}
%We now evaluate the beam misalignment behavior and its impact on BF performance under representative mmWave deployment settings. Unless otherwise specified, the number of beams at the UE is fixed at $N_{\mathrm{beam}}^{\mathrm{UE}} = 4$, the SSB burst set periodicity is $\tau_{\mathrm{SS}} = 20$~ms, and the SS burst duration is $T_{\mathrm{SS}} = 5$~ms. The beam processing delay (accounting for measurement, reporting, and beam update) is set to $T_{\mathrm{proc}} = 1$~ms. These parameters are selected in accordance with typical 3GPP NR configurations. The results capture the effects of BS beam count ($N_\mathrm{beam}^\mathrm{BS}$), user mobility ($v$), ISD ($d_{\mathrm{ISD}}$) and system configurations (i.e. SCS) on metrics such as misalignment duration, misalignment fraction, and average BF gain.
We evaluate beam misalignment and its impact on beamforming performance under typical mmWave deployment settings. Unless stated otherwise, parameters are set to $N_{\mathrm{beam}}^{\mathrm{UE}} = 4$, $\tau_{\mathrm{SS}}\!\!=\!\!20$ms, $T_{\mathrm{SS}}\!\!=\!\!5$ms, and $T_{\mathrm{proc}}\!\!=\!\!1$ms. The analysis explores the influence of BS beam count ($N_\mathrm{beam}^\mathrm{BS}$), user speed ($v$), inter-site distance ($d_{\mathrm{ISD}}$), and subcarrier spacing (SCS) on misalignment duration, misalignment fraction, and average beamforming gain.

\begin{figure}
    \centering
    \includegraphics[width=8cm]{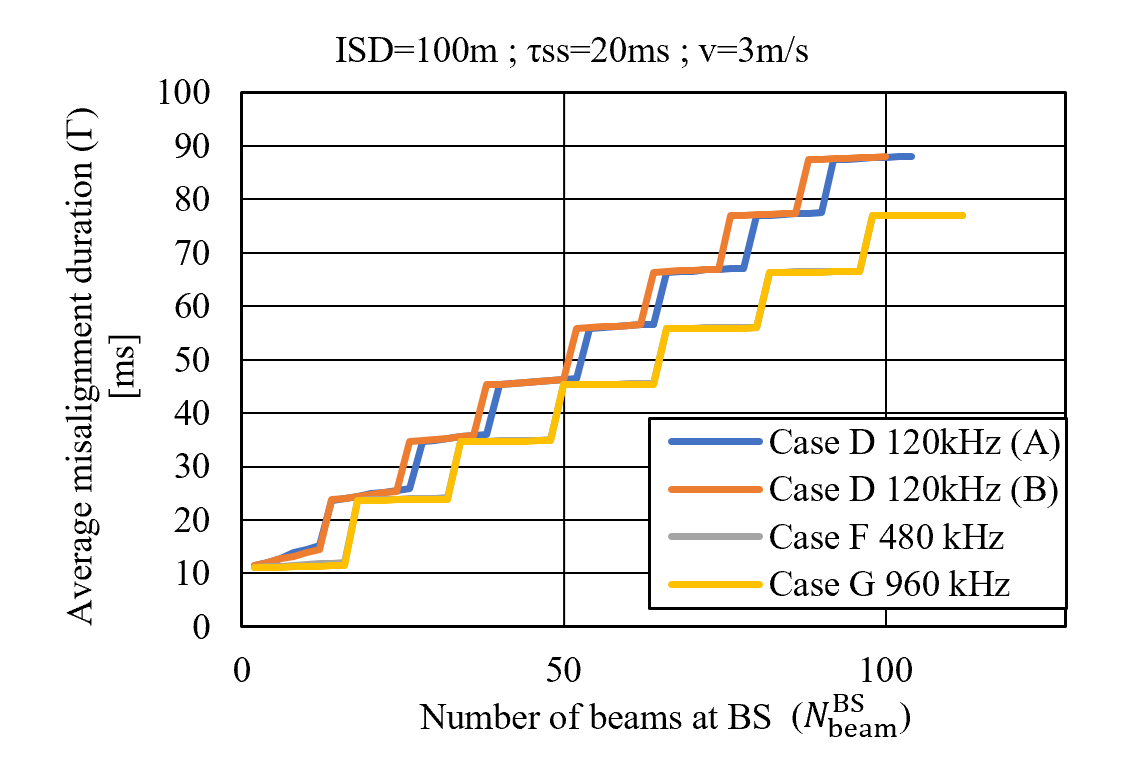}
    \caption{Average beam misalignment duration $\Gamma$ versus the number of BS beams $N_\mathrm{beam}^\mathrm{BS}$ for different mmWave cell configurations.}
    \label{fig:average_mis_duration}
\end{figure}

Fig. \ref{fig:average_mis_duration} shows the overall average beam misalignment duration ($\Gamma$), given by (\ref{eq:gamma_total_duration}), for various mmWave deployment scenarios (see Table \ref{tab:max_ssb_table}). As expected, $\Gamma$ increases approximately linearly with the number of BS beams, driven by the increase in total beam sweeping combinations $N_\mathrm{beam}^\mathrm{BS} \cdot N_\mathrm{beam}^\mathrm{UE}$. The stepwise shape of the curves stems from the discrete SSB sweeping behaviour illustrated in Fig.\ref{fig:sweep_duration}. Configurations with higher subcarrier spacing (e.g., Cases F and G) yield shorter misalignment durations due to their smaller symbol periods, which enable denser SSB packing and faster sweeping cycles. Differences between TDD patterns (Cases D-(a) vs D-(b)) and between Cases F and G are minor, indicating that the number of SSBs per SS burst set ($L_\mathrm{SS}$) is a dominant factor.

\begin{figure}
    \centering
    \includegraphics[width=8cm]{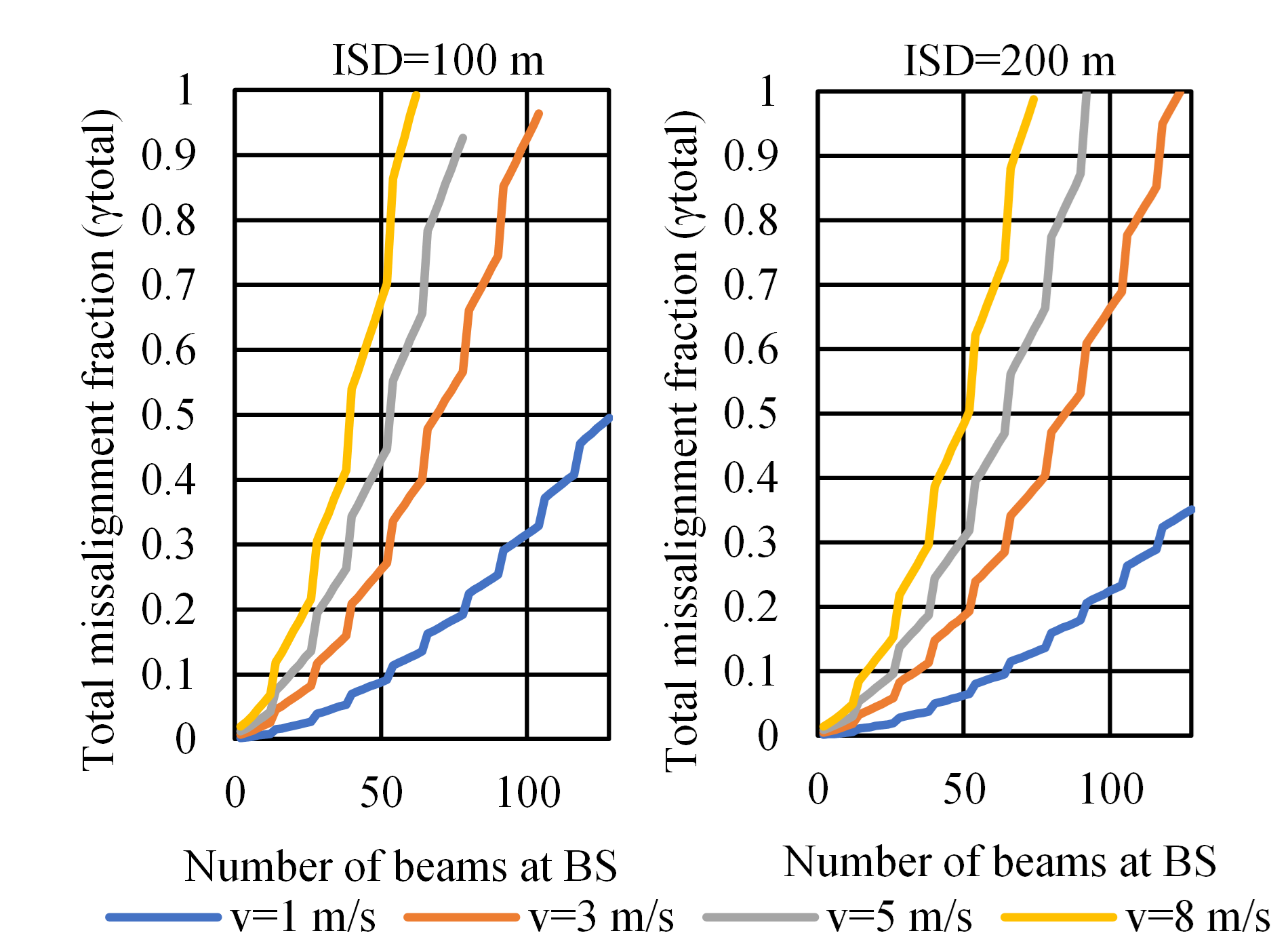}
    \caption{Total misalignment fraction against number of beams at the BS ($N_{\mathrm{beam}}^{\mathrm{BS}}$) for different UE speeds ($v$). Case D ($\Delta f=120$ kHz), $\tau_{\mathrm{SS}}=20$ ms. (left) $d_{\mathrm{ISD}}=100$m (right) $d_{\mathrm{ISD}}=200$m.  }
    \label{fig:mis_fraction}
\end{figure}

Fig. \ref{fig:mis_fraction} shows the total misalignment fraction $\gamma_\mathrm{total}$, defined in \eqref{eq:gamma_total}, against the number of BS beams ($N_\mathrm{beam}^\mathrm{BS}$) for two ISDs ($d_{\mathrm{ISD}}= \{ 100,200 \}$m) and various UE speeds. As expected, both the number of beams and UE speed significantly affect the misalignment fraction, with faster-moving UEs and denser beam configurations leading to a higher proportion of misalignment time. In contrast, the effect of ISD is modest, as its influence on the misalignment rate $\beta^{(i)}$ is through the square root of $\lambda$ in \eqref{eq:prob_beam_missalign}. For high UE speeds (e.g., $v=8$~m/s), the system becomes misaligned nearly 100\% of the time when using very narrow beams (e.g., $N_\mathrm{beam}^\mathrm{BS} > 80$), suggesting beam tracking mechanisms must adapt more rapidly, e.g. by reducing the SS burst set periodicity $\tau_{SS}$ (set to 20 ms in this scenario).

\begin{figure}
    \centering
    \includegraphics[width=8cm]{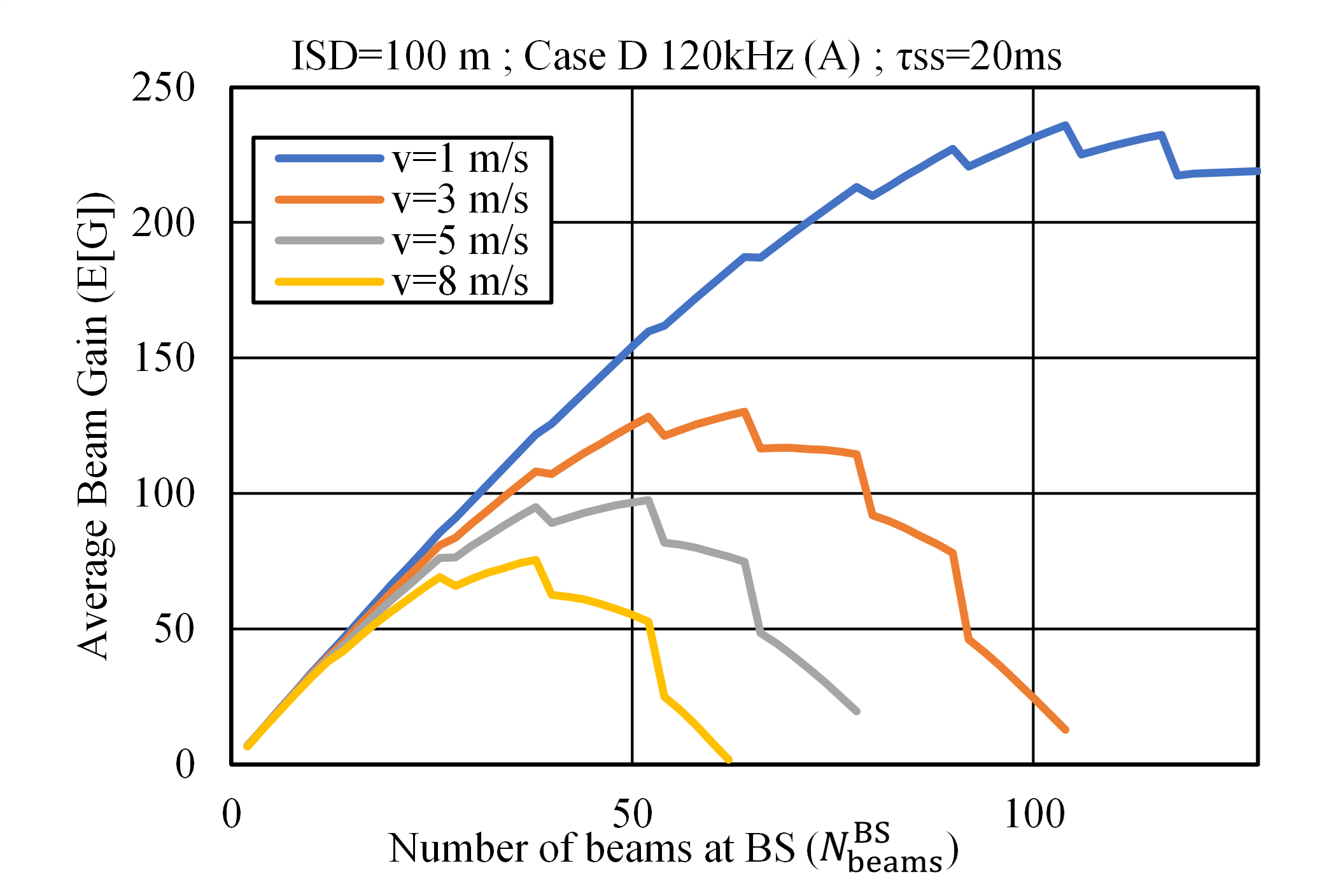}
    \caption{Average BF gain $\mathbb{E}[G]$ as a function of the number of BS beams ($N_\mathrm{beam}^\mathrm{BS}$) for different UE speeds $v$. Case D, 120~kHz subcarrier spacing (Pattern A), $\tau_\mathrm{SS} = 20$~ms, $d_{\mathrm{ISD}}$~m.}
    \label{fig:average_gain}
\end{figure}

Fig.~\ref{fig:average_gain} shows the average BF gain $\mathbb{E}[G]$, given by:
\begin{equation}
\begin{aligned}
    \mathbb{E}[G] = (1\!-\!\eta_\mathrm{OH}) \smashoperator{\prod_{i\in \{\mathrm{BS,UE}\}}} \left( (1\!-\!\gamma_{(i)}) \cdot G_{\mathrm{main}}^{(i)} \!+\!\gamma_{(i)}\cdot G_{\mathrm{side}}^{(i)}\right),
\end{aligned}
\end{equation}
where $\gamma_{\mathrm{BS}}$ and $\gamma_{\mathrm{UE}}$ are the misalignment fractions defined in~\eqref{eq:gamma_BS}, and $\eta_\mathrm{OH}$ denotes the SSB overhead, expressed as the ratio of DL symbols used for SSBs over the total number of DL symbols. As expected, increasing the number of BS beams ($N_{\mathrm{beam}}^{\mathrm{BS}}$) initially improves the average gain due to finer directional resolution. However, beyond a certain point, the gain starts to decline as increased misalignment due to UE mobility outweighs the benefits of narrower beams. This turning point occurs at lower BS beam counts for higher UE speeds, indicating reduced robustness to misalignment and tighter constraints on beam granularity under high mobility.

\section{Conclusion}
\label{sec:conclusion}
We presented a tractable analytical model to assess beam misalignment in mmWave systems under analog beamforming. By incorporating realistic NR configurations, including SSB sweeping behaviour, TDD frame timing, and user mobility, we quantified the long-term average impact of mutual beam misalignment between the BS and UE. Our analysis included expressions for the expected misalignment duration, the fraction of time the system operates in a misaligned state, and the resulting degradation in average beamforming gain.
The results highlight the sensitivity of misalignment performance to UE speed and beam count, while showing that the impact of inter-site distance is relatively limited. Narrow beams increase the risk of misalignment, especially under mobility, calling for tighter control of SSB periodicity and sweeping strategies. The framework and insights presented in this work provide useful guidelines for designing and optimizing beam management in mmWave and beyond, where maintaining alignment is crucial for achieving high directional gain and reliable connectivity. 

% or

\bibliographystyle{IEEEtran}
\bibliography{IEEEabrv,references}

\appendices
\section{TDD framestructure harmonization}
\label{appA}
The subcarrier spacing in 3GPP NR (SCS) takes the form $\Delta f=15\cdot2^\mu$ kHz, with numerology $\mu=3$ for 120kHz (case D), $\mu=4$ for 480kHz (case F) and $\mu=5$ for 960kHz (case G). Accordingly, the slot duration is given by $T_{\mathrm{slot}}=1/2^\mu$ ms. The proposed harmonized TDD framestructure by CEPT is given in \cite{ecc_rec_2003}, considering two different patterns, (a) and (b), which we will address in the following subsections.

\subsection{TDD Pattern (a)}
For pattern (a), the slot pattern periodicity is $T_{\mathrm{TDD}}^{(a)}=2.5$ ms, and the slot pattern duration (in slots) is given by:
\begin{equation}
N_{\mathrm{TDD}}^{(a),\mu} = 2.5 / T_{\mathrm{slot}} = 2.5 \cdot 2^\mu = 5 \cdot 2^{\mu-1} .
\end{equation}
For a slot duration of $T_{\mathrm{slot}}=0.5$ ms (i.e. $\mu=1$), the framestructure for pattern (a) follows a repeating pattern of \texttt{DDDSU}, where \texttt{D} indicates downlink (DL) slots, \texttt{S} indicates the special slot, and \texttt{U} indicates the uplink (UL) slots. The configuration of special slot \texttt{S} follows the pattern DL:GP:UL, where the number of DL, Guard Period (GP), and UL symbols are given as 10, 2, and 2, respectively.
For subcarrier spacings $\mu>0$, we propose a proportional TDD pattern scaling which ensures alignment of DL, S, and UL slots across different numerologies $\mu$. Accordingly, the number of consecutive DL slots in pattern (a) and numerology $\mu$ is given by: 
\begin{equation}
N_{\mathrm{TDD,DL}}^{(a),\mu} = 2^{\mu+1} - 1 .
\end{equation}

We keep the number of special slots equal to 1, and with the same DL:GP:UL configuration as for $\mu=1$, therefore:
\begin{equation}
N_{\mathrm{TDD,S}}^{(a),\mu} = 1 .
\end{equation}

Finally, the number of consecutive UL slots in pattern (a) and numerology $\mu$ is given by:
\begin{equation}
N_{\mathrm{TDD,UL}}^{(a),\mu} = 2^{\mu-1} .
\end{equation}

Table \ref{tab:tdd_ul_dl_pattern} shows, for a range of numerologies $\mu$, the number of DL, S and UL slots in a framestructure following the pattern option (a).

\begin{table}[htbp]
\centering
\caption{TDD UL/DL Pattern Option (A) for different numerologies.}
\label{tab:tdd_ul_dl_pattern}
\resizebox{\columnwidth}{!}{%
\begin{tabular}{|c|c|c|c|c|c|c|}
\hline
\textbf{$\mu$} & \textbf{SCS} & \textbf{DL slots} & \textbf{S slots} & \textbf{UL slots} & \textbf{Total} & $\textbf{Slot Duration}$ \\
 & $(\Delta f)$ & $N_{\mathrm{TDD,DL}}^{(a),\mu}$ & $N_{\mathrm{TDD,S}}^{(a),\mu}$ & $N_{\mathrm{TDD,UL}}^{(a),\mu}$ & \textbf{slots} & $(T_{\mathrm{slot}})$\\
 & [kHz] & [slots] & [slots] & [slots] & [slots] & [ms/slot] \\
\hline
1  & 30   & 3  & 1  & 1  & 5  & 0.5 \\
2  & 60   & 7  & 1  & 2  & 10 & 0.25 \\
3  & 120  & 15 & 1  & 4  & 20 & 0.125 \\
4  & 240  & 31 & 1  & 8  & 40 & 0.0625 \\
5  & 480  & 63 & 1  & 16 & 80 & 0.03125 \\
6  & 960  & 127 & 1  & 32 & 160 & 0.015625 \\
\hline
\end{tabular}
}
\end{table}

\subsection{TDD Pattern (b)}
For Pattern (b), the slot pattern periodicity is $T_{tdd}^{(b)} = 5$ ms, and the slot pattern duration (in slots) is given by: 
\begin{equation} 
N_{\mathrm{TDD}}^{(b),\mu} = 5 / T_{\mathrm{slot}} = 5 \cdot 2^\mu . 
\end{equation} 

For a slot duration of $T_{\mathrm{slot}} = 0.5$ ms (i.e., $\mu = 1$), the frame structure for Pattern (b) follows a repeating pattern of \texttt{DDDSUUDDD}, where \texttt{D}, \texttt{S} and \texttt{U} indicates DL, S and UL slots respectively. The configuration of \texttt{S} follows the pattern DL:GP:UL, where the number of DL, Guard Period (GP), and UL symbols are given as 6, 4, and 4, respectively, or 4,6, and 4 \cite{ecc_rec_2003}.

For subcarrier spacings $\mu > 0$, we propose a proportional TDD pattern scaling which ensures alignment of DL, S, and UL slots across different numerologies $\mu$. Accordingly, the number of consecutive DL slots in Pattern (b) and numerology $\mu$ is given by:
\begin{equation} 
N_{\mathrm{TDD,DL}}^{(b),\mu} = 2^{\mu+2} - 1 . 
\end{equation}

We keep the number of special slots equal to 1, and with the same DL:GP:UL configuration as for $\mu = 1$, therefore: 
\begin{equation} 
N_{\mathrm{TDD,S}}^{(b),\mu} = 1 . 
\end{equation}

Finally, the number of consecutive UL slots in Pattern (b) and numerology $\mu$ is given by: 
\begin{equation} 
N_{\mathrm{TDD,UL}}^{(b),\mu} = 2^{\mu}. 
\end{equation}

Table \ref{tab:tdd_ul_dl_pattern_b} shows, for a range of numerologies $\mu$, the number of DL, S, and UL slots in a frame structure following Pattern (b).

\begin{table}[htbp] 
\centering 
\caption{TDD UL/DL Pattern Option (B) for different numerologies.} \label{tab:tdd_ul_dl_pattern_b} 
\resizebox{\columnwidth}{!}{%
\begin{tabular}{|c|c|c|c|c|c|c|} 
\hline 
\textbf{$\mu$} & \textbf{SCS} & \textbf{DL slots} & \textbf{S slots} & \textbf{UL slots} & \textbf{Total} & $\textbf{Slot Duration}$ \\ 
& $(\Delta f)$ & $N_{\mathrm{TDD,DL}}^{(b),\mu}$ & $N_{\mathrm{TDD,S}}^{(b),\mu}$ & $N_{\mathrm{TDD,UL}}^{(b),\mu}$ & \textbf{slots} & $(T_{\mathrm{slot}})$\\
& [kHz] & [slots] & [slots] & [slots] & [slots] & [ms/slot] \\
\hline 
1 & 30 & 7 & 1 & 2 & 10 & 0.5 \\
2 & 60 & 15 & 1 & 4 & 20 & 0.25 \\
3 & 120 & 31 & 1 & 8 & 40 & 0.125 \\
4 & 240 & 63 & 1 & 16 & 80 & 0.0625 \\
5 & 480 & 127 & 1 & 32 & 160 & 0.03125 \\
6 & 960 & 255 & 1 & 64 & 320 & 0.015625 \\
\hline 
\end{tabular} 
} 
\end{table}

\subsection{TDD slot indexing}
The time-domain slot indexing for a given pattern $p=\{a,b\}$, with $n=\{0,1,2,...\}$ can be written as:
\begin{equation}
\label{eq:set_DL_slots_frame_a}
\mathcal{N}_{\mathrm{DL}}^{(p),\mu} = \{0, 1, 2, \dots, N_{\mathrm{TDD,DL}}^{(p),\mu} - 1\} + N_{\mathrm{TDD}}^{(p),\mu} \cdot n
\end{equation}
for the DL slots,
\begin{equation}
\mathcal{N}_{\mathrm{S}}^{(p),\mu} = N_{\mathrm{TDD,DL}}^{(p),\mu} + N_{\mathrm{TDD}}^{(p),\mu} \cdot n
\end{equation}
for the special slot, and
\begin{equation}
\begin{aligned}
\mathcal{N}_{\mathrm{UL}}^{(p),\mu} & = \{N_{\mathrm{TDD,DL}}^{(p),\mu}  +  N_{\mathrm{TDD,S}}^{(p),\mu}, \dots, N_{\mathrm{TDD}}^{(p),\mu} - 1\} \\ & + N_{\mathrm{TDD}}^{(p),\mu} \cdot n    
\end{aligned}
\end{equation}
for the UL slots.

\section{SSB transmission starting symbols}
\label{appB}
Next we provide the derivation of the starting symbols (referred to the start of a NR radio frame) for cases D ($\mu=3$), F ($\mu=4$) and G ($\mu=5$) in \cite{3gpp_ts_38213}.

\subsection{Case D - 120kHz - Framestructure-agnostic}
\label{subsec:case_D_frame_agnostic}
According to \cite{3gpp_ts_38213}, the set containing the starting symbols for each SSB is given by the following relationship:
\begin{equation}
\begin{aligned}
\mathcal{L}_{\mathrm{SSB}}^D = \{4, 8, 16, 20\} + 28 \cdot n,\\
n \in \{0, 1, 2, 3, 5, 6, 7, 8,\\
\phantom{n \in \{}10, 11, 12, 13, 15, 16, 17, 18\} \\
\end{aligned}
\end{equation}
This results in a total number of SSBs per SS burst set of:
\begin{equation}
L_{\mathrm{SSB}}^D = \left| \mathcal{L}_{\mathrm{SSB}}^D \right| = 64.    
\end{equation}
The $i$-th SSB symbol start $l_{\mathrm{SSB}}^D(i)$ can be drawn from the set $\mathcal{L}_{\mathrm{SSB}}^D$, as $l_{\mathrm{SSB}}^D(i)\in \mathcal{L}_{\mathrm{SSB}}^D$ with $i=1, 2,...\left| \mathcal{L}_{\mathrm{SSB}}^D \right|$. 

\subsection{Case D - 120kHz - TDD Option (a)}
The set of effective slots that can carry SSBs is defined by the intersection of the total set, as described in section \ref{subsec:case_D_frame_agnostic}, with the set of avaliable DL slots as described by (\ref{eq:set_DL_slots_frame_a}) with $p=\{a\}$ and $\mu=3$:
\begin{equation}
\mathcal{L}_{\mathrm{SSB,eff}}^{(a),D} = \mathcal{L}_{\mathrm{SSB}}^D\pmod{14} \;\cap\; \mathcal{N}_{\mathrm{DL}}^{(a),3}.
\end{equation}

This results in the set of effective SSB start symbols (only DL slots allowed):

\begin{equation}
\mathcal{L}_{\mathrm{SSB,eff}}^{(a),D} =
\begin{cases}
\{4, 8, 16, 20\} + 28 \cdot n, & n \in \{0, 1, 2, 3, 5, 6, 10,\\
 &  11, 12, 13, 15, 16, 18\} \\
\{4, 8\} + 28 \cdot n,         & n \in \{7, 17\}
\end{cases}
\end{equation}
yielding:
\begin{equation}
{L}_{\mathrm{SSB,eff}}^{(a),D}=\left| \mathcal{L}_{\mathrm{SSB,eff}}^{(a),D} \right| = 52.
\end{equation}

If both DL and F slots are allowed, the effective start symbols become:
\begin{equation}
\begin{aligned}
\mathcal{L}_{\mathrm{SSB,eff}}^{(a),D} &= \{4, 8, 16, 20\} + 28 \cdot n, \\
n & \in \{0, 1, 2, 3, 5, 6, 7, 10, 11, 12, \\ 
  & \phantom{\in \{} 13, 15, 16, 17, 18\},   
\end{aligned}
\end{equation}
resulting in:
\begin{equation}
\left| \mathcal{L}_{\mathrm{SSB,eff}}^{(a),D} \right| = 56.
\end{equation}

\subsection{Case D - 120kHz - TDD Option (b)}
The effective slots that can carry SSBs are given by the intersection of sets:
\begin{equation}
\mathcal{L}_{\mathrm{SSB,eff}}^{(b)} = \mathcal{L}_{\mathrm{SSB}}^D\pmod{14} \cap \mathcal{N}_{\mathrm{DL}}^{(b),3}.
\end{equation}

The set of effective SSB start symbols is:
\begin{equation}
\mathcal{L}_{\mathrm{SSB,eff}}^{(b),D} =
\begin{cases}
 \{4, 8, 16, 20\} + 28 \cdot n, & n \in \{0, 1, 2, 3, 5, 6,\\
 & 7, 8, 10, 11, 12, 13\} \\
 \{4, 8\} + 28 \cdot n, & n = 15 
\end{cases}
\end{equation}

This leads to:
\begin{equation}
{L}_{\mathrm{SSB,eff}}^{(b),D} = \left| \mathcal{L}_{\mathrm{SSB,eff}}^{(b),D} \right| = 50.
\end{equation}

If DL and F slots are allowed, the effective set of SSB start symbols becomes:
\begin{equation}
\begin{aligned}
\mathcal{L}_{\mathrm{SSB,eff}}^{(b),D} &= \{4, 8, 16, 20\} + 28 \cdot n, \\
& n \in \{0, 1, 2, 3, 5, 6, 7, 8, 10, 11, 12, 13, 15\},   
\end{aligned}
\end{equation}
This results in:
\begin{equation}
\left| \mathcal{L}_{\mathrm{SSB,eff}}^{(b),D} \right| = 52 .
\end{equation}

\subsection{Case F - 480kHz - Framestructure-agnostic}
For case F, i.e. $\mu=5$ and $\Delta f=480$ kHz, the starting symbols for each SSB within an SS burst set is given by \cite{3gpp_ts_38213}: 
\begin{equation}
\begin{aligned}
\mathcal{L}_{\mathrm{SSB}}^{F} = \{2, 9\} + 14 \cdot n, n \in \{ 0, 1, 2, \dots, 31 \},
\end{aligned}
\end{equation}
which results in:
\begin{equation}
\label{eq:ssb_symb_start_F_agnostic}
\begin{aligned}
\mathcal{L}_{\mathrm{SSB}}^{F} &= \{2, 9, 16, 23, 30, 37, 44, 51, 58, 65, 72, 79, 86, 93, \\
& \phantom{\in \{} 100, 107, 114, 121, 128, 135, 142, 149, 156, 163,\\
& \phantom{\in \{}  170, 177, 184, 191, 198, 205, 212, 219, 226, 296,\\
& \phantom{\in \{} 233, 240, 247, 254, 261, 268, 275, 282, 289, 366, \\
& \phantom{\in \{} 303, 310, 317, 324, 331, 338, 345, 352, 359, 436, \\
& \phantom{\in \{} 373, 380, 387, 394, 401, 408, 415, 422, 429,  443 \}. \\ 
\end{aligned}
\end{equation}
Accordingly, the total number of SSBs per SS burst set is:
\begin{equation}
\begin{aligned}
{L}_{\mathrm{SSB}}^{F} &= \left| \mathcal{L}_{\mathrm{SSB}}^{F} \right| = 64.
\end{aligned}
\end{equation}

\subsection{Case F - 480kHz - TDD Option (a)}
\label{sub:cese_F_480_TDD_A}
When adopting TDD option (a), the starting symbol set $\mathcal{L}_{\mathrm{SSB}}^{F}$ defined in (\ref{eq:ssb_symb_start_F_agnostic}) is totally aligned with the set of DL symbols in TDD option (a),  $\mathcal{N}_{\mathrm{DL}}^{(a),5}$, that is, $\mathcal{L}_{\mathrm{SSB}}^F\pmod{14} \subset \mathcal{N}_{\mathrm{DL}}^{(a),5}$, therefore
\begin{equation}
\begin{aligned}
{L}_{\mathrm{SSB,eff}}^{(a),F} &= \left| \mathcal{L}_{\mathrm{SSB}}^{F} \right| = 64.
\end{aligned}
\end{equation}
\subsection{Case F - 480kHz - TDD Option (b)}
Similarly, when adopting TDD option (b), the starting symbol set $\mathcal{L}_{\mathrm{SSB}}^{F}$ defined in (\ref{eq:ssb_symb_start_F_agnostic}) is totally aligned with the set of DL symbols in TDD option (b),  $\mathcal{N}_{\mathrm{DL}}^{(b),5}$, that is, $\mathcal{L}_{\mathrm{SSB}}^F\pmod{14} \subset \mathcal{N}_{\mathrm{DL}}^{(b),5}$, therefore
\begin{equation}
\begin{aligned}
{L}_{\mathrm{SSB,eff}}^{(b),F} &= \left| \mathcal{L}_{\mathrm{SSB}}^{F} \right| = 64.
\end{aligned}
\end{equation}

\subsection{Case G - 960kHz - Framestructure-agnostic}
With case G, considering $\Delta f=960$ kHz ($\mu=6$), the starting symbols for the SSB transmissions are defined by \cite{3gpp_ts_38213}:
\begin{equation}
\mathcal{L}_{\mathrm{SSB}}^{G} = \{2, 9\} + 14 \cdot n, \quad n \in \{ 0, 1, 2, \dots, 31 \},
\end{equation}
which result in
\begin{equation}
\begin{aligned}
\mathcal{L}_{\mathrm{SSB}}^{G} & = \{2, 9, 16, 23, 30, 37, 44, 51, 58, 65, 72, 79, 86, 93, \\
& \phantom{= \{,} 100, 107, 114, 121, 128, 135, 142, 149, 156, 163, \\
& \phantom{= \{,} 170, 177, 184, 191, 198, 205, 212, 219, 226, 233, \\
& \phantom{= \{,} 240, 247, 254, 261, 268, 275, 282, 289, 296, 303, \\
& \phantom{= \{,} 310, 317, 324, 331, 338, 345, 352, 359, 366, 373, \\
& \phantom{= \{,} 380, 387, 394, 401, 408, 415, 422, 429, 436, 443 \}  .
\end{aligned}
\end{equation}

Then, the total number of supported SSBs per SS burst set is given by:
\begin{equation}
{L}_{\mathrm{SSB}}^{G} = \left| \mathcal{L}_{\mathrm{SSB}}^{G} \right| = 64
\end{equation}

\subsection{Case G - 960kHz - TDD Option (a)}
Similar to case F, see section \ref{sub:cese_F_480_TDD_A}, set  $\mathcal{L}_{\mathrm{SSB}}^{G}$ fully aligns with the DL symbols in TDD framestructure (a), i.e. $\mathcal{L}_{\mathrm{SSB}}^G\pmod{14} \subset \mathcal{N}_{\mathrm{DL}}^{(a),6}$, therefore
\begin{equation}
\begin{aligned}
{L}_{\mathrm{SSB,eff}}^{(a),G} &= \left| \mathcal{L}_{\mathrm{SSB}}^{G} \right| = 64.
\end{aligned}
\end{equation}

\subsection{Case G - 960kHz - TDD Option (b)}
Equally, set  $\mathcal{L}_{\mathrm{SSB}}^{G}$ fully aligns with the DL symbols in TDD framestructure (b), i.e. $\mathcal{L}_{\mathrm{SSB}}^G\pmod{14} \subset \mathcal{N}_{\mathrm{DL}}^{(b),6}$, therefore
\begin{equation}
\begin{aligned}
{L}_{\mathrm{SSB,eff}}^{(b),G} &= \left| \mathcal{L}_{\mathrm{SSB}}^{G} \right| = 64.
\end{aligned}
\end{equation}

\section{SSB sweep time calculation}
\label{appC}
The aim is to allocate an SSB request of $N_{\mathrm{SSB}}^{\mathrm{req}}$ SSBs. The available maximum possible allocations within one SS burst set is ${L}_{\mathrm{SSB,eff}}^{(p),\mu}=|\mathcal{L}_{\mathrm{SSB,eff}}^{(p),\mu}|$, as established in appendix \ref{appB}, with $p=\{a,b\}$ representing the TDD framestructure options (a) and (b), and $\mu$ the numerology indicating the subcarrier spacing relevant to cases D, F and G.

If $N_{\mathrm{SSB}}^{\mathrm{req}} >{L}_{\mathrm{SSB,eff}}^{(p),\mu}$, then the residual SSBs are allocated during the following SS burst period (after a time $T_{\mathrm{SS}}$).

The number of SSBs per slot (SS burst size) is given by $N_{\mathrm{SSB}}^{\mathrm{slot}} = \{1, 2\}$ SSBs \cite{3gpp_ts_38213}. We will assume hereon that $N_{\mathrm{SSB}}^{\mathrm{slot}} = 2$.

\begin{figure*}[!htb]
% ensure that we have normalsize text
%\normalsize
% Store the current equation number.
\setcounter{MYtempeqncnt}{\value{equation}}
% Set the equation number to one less than the one
% desired for the first equation here.
% The value here will have to changed if equations
% are added or removed prior to the place these
% equations are referenced in the main text.
\setcounter{equation}{54}
\begin{equation}
\label{eq:span1}
N_{\mathrm{sweep,r}}(n_{\mathrm{SS}}) = 
\begin{cases}
I_1(n_{\mathrm{SS}}) \cdot \left[ \Lambda(n_\mathrm{SS}) + I_2(n_{\mathrm{SS}}) \cdot \left( \frac{r(n_{\mathrm{SS}})}{N_{\mathrm{SSB}}^{\mathrm{slot}}} \cdot 14 - N_{\mathrm{symb}}^{-}(n_{\mathrm{SS}}) \right) \right], & \text{if } r(n_{\mathrm{SS}}) \ (\text{mod} \ 2) = 0 \\
I_1(n_{\mathrm{SS}}) \cdot \left[ \Lambda(n_\mathrm{SS}) + I_2(n_{\mathrm{SS}}) \cdot \left( \left\lfloor \frac{r(n_{\mathrm{SS}})}{N_{\mathrm{SSB}}^{\mathrm{slot}}} \right\rfloor \cdot 14 + N_{\mathrm{symb}}^{+}(n_{\mathrm{SS}}) \right) \right], & \text{if } r(n_{\mathrm{SS}}) \ (\text{mod} \ 2) \neq 0
\end{cases}
\end{equation}
with
\begin{equation*}
    \Lambda(n_\mathrm{SS})\triangleq (1 - I_2(n_{\mathrm{SS}})) \cdot \left( \frac{r(n_{\mathrm{SS}})}{N_{\mathrm{SSB}}^{\mathrm{slot}}} \cdot 14 + G_{\mathrm{SSB}}(n_{\mathrm{SS}}) \cdot 14 \right).
\end{equation*}
% Restore the current equation number.
\setcounter{equation}{\value{MYtempeqncnt}}
% IEEE uses as a separator
\hrulefill
% The spacer can be tweaked to stop underfull vboxes.
%\vspace*{4pt}
\end{figure*}

The SS burst set can be divided into a number of $N_{\mathrm{SS}}$ burst segments. Each segment is defined as a consecutive number of slots, each containing $N_{\mathrm{SSB}}^{\mathrm{slot}}$ SSBs. Let the burst segments be indexed by $n_{\mathrm{SS}} = 1, 2, \ldots, N_{\mathrm{SS}}$. Burst segments are configuration-specific, each with a specific SSB capacity $c_{\mathrm{SSB}}(n_{\mathrm{SS}})$. Since SSB allocations are assumed to be sequential (i.e., filling each segment to capacity before moving to the next segment), we define the cumulative capacity function at segment $n_{\mathrm{SS}}$ as:
\begin{equation}
C(n_{\mathrm{SS}}) = 
\begin{cases}
0, & n_{\mathrm{SS}} = 0 \\
\sum_{k=1}^{n_{\mathrm{SS}}} c_{\mathrm{SSB}}(k), & 0 < n_{\mathrm{SS}} \leq N_{\mathrm{SS}}
\end{cases}\phantom{===}.
\end{equation}

The total capacity per SS burst set is then:
\begin{equation}
C(N_{\mathrm{SS}}) = \sum_{k=1}^{N_{\mathrm{SS}}} c_{\mathrm{SSB}}(k)\phantom{=}.
\end{equation}

We define the number of completely allocated SS burst sets $n_\mathrm{c}$ (not including the last SS burst set, whether incomplete or complete) as:

\begin{equation}
n_\mathrm{c} = \left\lceil \frac{N_{\mathrm{SSB}}^{\mathrm{req}}}{C(N_{\mathrm{SS}})} \right\rceil - 1 \phantom{=}.
\end{equation}

This accounts for the completely allocated SS burst sets and the still remaining SSBs to be allocated in the next SS burst set (important when calculating the total SSB sweep time duration).

The total remaining SSBs to be allocated in the last burst set, $n_\mathrm{r}$, can be defined as:
\begin{equation}
n_\mathrm{r} = N_{\mathrm{SSB}}^{\mathrm{req}} - n_\mathrm{c}.
\end{equation}

The residual SSBs to be allocated in each burst segment of the last SS burst set can be computed as:
\begin{equation}
r(n_{\mathrm{SS}}) = \min \left( c_{\mathrm{SSB}}(n_{\mathrm{SS}}), \max(0, n_\mathrm{r} - C(n_{\mathrm{SS}} - 1)) \right),
\end{equation}
where $n_\mathrm{r} - C(n_{\mathrm{SS}} - 1)$ represents the remaining SSBs after filling previous segments, $\max(0, \cdot)$ ensures a non-negative allocation, and $\min(c_{\mathrm{SSB}}(n_{\mathrm{SS}}), \cdot)$ limits the allocation to the segment's capacity. For mathematical convenience, we define $r(0) = 0$.

Next we define the following indicator functions for mathematical convenience. Firstly, $I_1(n_{\mathrm{SS}})$ reflects whether segment $n_{\mathrm{SS}}$ has any residual allocations, equaling 1 if it does, or 0 otherwise:
\begin{equation}
I_1(n_{\mathrm{SS}}) = 
\begin{cases}
\left\lfloor \frac{\sign(r(n_{\mathrm{SS}})) + 1}{2} \right\rfloor, & n_{\mathrm{SS}} \leq N_{\mathrm{SS}} \\
0, & n_{\mathrm{SS}} \geq N_{\mathrm{SS}}
\end{cases}\phantom{=}.
\end{equation}

Secondly, $I_2(n_{\mathrm{SS}})$ identifies the last segment with residual allocations:
\begin{equation}
I_2(n_{\mathrm{SS}}) = I_1(n_{\mathrm{SS}}) - I_1(n_{\mathrm{SS}} + 1).
\end{equation}

Similarly to burst segments, we define $N_{\mathrm{SS}}$ SS burst gaps, immediately following the burst segments, as consecutive slots that do not contain SSBs (e.g., non-allocated slots or special and UL slots). Burst gaps are also configuration-specific, each with a specific SS gap duration (in slots) of $G_{\mathrm{SS}}(n_{\mathrm{SS}})$ with $n_{\mathrm{SS}} = 1, \ldots, N_{\mathrm{SS}}$. Also, $G_{\mathrm{SS}}(0) = 0$.

The time required to sweep all allocated SSBs is the sum of the time needed to sweep complete SS burst sets ($N_{\mathrm{SSB},c}$), plus the time needed to sweep residual SS burst sets ($N_{\mathrm{SSB},r}$), in symbols:
\begin{equation}
\label{eq:sweep_symbols}
N_{\mathrm{sweep}} = N_{\mathrm{sweep,c}} + N_{\mathrm{sweep,r}}
\end{equation}

The time needed to sweep complete SS bursts ($N_{\mathrm{sweep,c}}$), in symbols, is given by:
\begin{equation}
N_{\mathrm{sweep,c}} = n_\mathrm{c} \cdot \left( \frac{\tau_{\mathrm{SS}} \cdot 14}{T_{\mathrm{slot}}} \right) = \left( \left\lfloor \frac{C(N_{\mathrm{SS}})}{N_{\mathrm{SSB}}^{\mathrm{req}}} \right\rfloor - 1 \right) \cdot \tau_{\mathrm{SS}} \cdot 2^{\mu}
\end{equation}
where $n_\mathrm{c}$ is the number of completed SS burst set allocations, $\tau_{\mathrm{SS}}$ is the SS burst set periodicity (in ms), 14 symbols/slot, and $T_{\mathrm{slot}} = {1}/{2^{\mu}}$ is the slot duration (in ms).

The time needed to sweep the SSBs in the residual SS burst set ($N_{\mathrm{sweep,r}}$) depends on the allocated SSB resources over the SS burst segments and the possible SS gaps between them. With the aid of previously defined indicator functions, we have:
\begin{equation}
N_{\mathrm{sweep,r}} = \sum_{n_{\mathrm{SS}}=1}^{N_{\mathrm{SS}}} N_{\mathrm{sweep,r}}(n_{\mathrm{SS}}),
\end{equation}
with $N_{\mathrm{sweep,r}}$ given by equation (\ref{eq:span1}).
\setcounter{equation}{55}
In (\ref{eq:span1}) the number of remaining symbols in the slot after the last SSB allocation, $N_{\mathrm{symb}}^{-}(n_{\mathrm{SS}}) $ is defined as:
\begin{equation}
N_{\mathrm{symb}}^{-}(n_{\mathrm{SS}}) = 14 - (N_{\mathrm{start,last}}(n_{\mathrm{SS}}) + N_{\mathrm{SSB}}^{\mathrm{symb}}),
\end{equation}
where $N_{\mathrm{start,last}}(n_\mathrm{SS})$ depends on the particular case (i.e. $\mu$) and the slot index which carries the last SSB allocation (whether it is even or odd). In particular, for case D:
\begin{equation}
N_{\mathrm{start,last}}^{D}(n_{\mathrm{SS}}) =
\begin{cases}
8, & \text{if } n_{\mathrm{SSB,last slot}}(n_{\mathrm{SS}}) \ (\text{mod} \ 2) = 0 \\
6, & \text{if } n_{\mathrm{SSB,last slot}}(n_{\mathrm{SS}}) \ (\text{mod} \ 2) = 1
\end{cases}.
\end{equation}
For cases F and G:
\begin{equation}
N_{\mathrm{start,last}}^{\{F,G\}}(n_{\mathrm{SS}}) = 9.
\end{equation}

For $N_{\mathrm{symb}}^{+}(n_{\mathrm{SS}})$ in (\ref{eq:span1}), i.e. the symbol number corresponding to the last symbol in the last SSB allocation is defined as:
\begin{equation}
N_{\mathrm{symb}}^{+}(n_{\mathrm{SS}}) = N_{\mathrm{start,first}}(n_{\mathrm{SS}}) + N_{\mathrm{SSB}}^{\mathrm{symb}},
\end{equation}
where $N_{\mathrm{start,first}}(n_\mathrm{SS})$ also depends on the case ($\mu$) and the slot index which carries the last SSB allocation (whether it is even or odd). In particular, for case D:
\begin{equation}
N_{\mathrm{start,first}}^{D}(n_{\mathrm{SS}}) =
\begin{cases}
4, & \text{if } n_{\mathrm{SSB,last slot}}(n_{\mathrm{SS}}) \ (\text{mod} \ 2) = 0 \\
2, & \text{if } n_{\mathrm{SSB,last slot}}(n_{\mathrm{SS}}) \ (\text{mod} \ 2) = 1
\end{cases}.
\end{equation}
And for cases F and G:
\begin{equation}
N_{\mathrm{start,first}}^{\{F,G\}}(n_{\mathrm{SS}}) = 2 \quad \forall n_{\mathrm{SS}}.
\end{equation}

The slot index corresponding to the last SSB allocation in segment $n_{\mathrm{SS}}$ is given by:
\begin{equation}
\begin{aligned}
    n_{\mathrm{SSB,last slot}}(n_{\mathrm{SS}}) & = \left\lceil \frac{r(n_{\mathrm{SS}})}{N_{\mathrm{SSB}}^{\mathrm{slot}}} - 1 \right\rceil \\ 
    & + \sum_{k=1}^{n_{\mathrm{SS}}} \left( G_{\mathrm{SS}}(n_{\mathrm{SS}} - k) + \frac{r(n_{\mathrm{SS}} - k)}{N_{\mathrm{SSB}}^{\mathrm{slot}}} \right)
\end{aligned}
\end{equation}
where the summation accounts for slot allocations in previous segments and corresponding SS burst gaps.

Finally, the time needed to sweep all allocated SSBs can be derived from (\ref{eq:sweep_symbols}) as:
\begin{equation}
\label{eq:sweep_time}
\begin{aligned}
T_{\mathrm{sweep}} = & T_{\mathrm{sweep,c}} + T_{\mathrm{sweep,r}}= \\&(T_{\mathrm{slot}}/14)\cdot(N_{\mathrm{sweep,c}} + N_{\mathrm{sweep,r}}).    
\end{aligned}
\end{equation}


% % you can choose not to have a title for an appendix
% % if you want by leaving the argument blank
% \section{}

% \section*{Acknowledgment}
% This work has received funding from the European Union's Horizon Europe research and innovation programme under the Marie Skłodowska-Curie grant agreement No. 101073265 (EWOC). Views and opinions expressed are however those of the authors only and do not necessarily reflect those of the European Union. The European Union cannot be held responsible for them.

\ifCLASSOPTIONcaptionsoff
  \newpage
\fi

\end{document}